\documentclass[prb,twocolumn,10pt,nofootinbib]{revtex4-1}
\usepackage{graphicx,amssymb,amsmath,subfigure,hyperref}
\hypersetup{colorlinks=true, citecolor=blue,linkcolor=red}
\usepackage[usenames]{color}
\newcommand{\ket}[1]{\left|#1\right>}

\newcommand{\para}[1]{\left(#1\right)}

\newcommand{\sd}[0]{\ \ \ }

\begin{document} 
\title{UV-IR transmutation for hybrid realizations of $\mathbb{Z}_k$ parafermion systems }
\author{Abolhassan Vaezi, Eun-Ah Kim}
\affiliation{Department of Physics, Cornell University, Ithaca, New York 14853, USA}


\begin{abstract}
Recent experiments brought Majorana particles closer to reality and raised interest in  
realizing $\mathbb{Z}_k$ parafermions for general value of $k$ in hybrid structures. Of particular
interest is the prospect of realizing a two dimensional system with the edge states described by $\mathbb{Z}_k$ parafermion conformal field theory (CFT) by coupling one dimensional hybrid systems. However in order to understand the effective perturbation due to local inter-wire coupling,
it is crucial to relate short distance ultra-violet (UV) degrees of freedom to primary fields of the CFT describing long distance infra-red (IR) physics. Here we consider two recent proposals for realizing a one dimensional system described by $\mathbb{Z}_k$ parafermion CFT upon tuning to quantum critical points: a parafermion chain and a pair of $\nu=2/k$ quantum Hall chiral edge states under uniform backscattering and pairing, where the latter is represented by self-dual sine-Gordon model. We explicitly derive the relation between distinct UV degrees of freedom relevant for each proposals and the primary fields of $\mathbb{Z}_3$ parafermion CFT. Our results point to a marked difference in the effect of local coupling between wires represented by each proposals, regarding robustness of the edge states in the resulting two dimensional systems.
\end{abstract}

\maketitle
\section{Introduction}
Potential realization of exotic Majorana particles which are their own particle-hole conjugates localized  at the end of semiconducting wires\cite{Delft} raised broad interest in these exotic particles. In addition to the proposals\cite{wire-proposal1,wire-proposal2} that led to the semiconducting wire setup of Ref.~\cite{Delft},  various proposals for realizing vortex bound states of chiral triplet superconductor with $p_x+ip_y$ pairing structure\cite{RG,Fu-Kane,QHZ} are under active pursuit as ways of realizing localized majorana zero modes.
A rather different state of majorana operator extended in one spatial dimension (1D) can be realized as a primary field of Ising conformal field theory (CFT) describing gapless edge excitations of filling factor $\nu=5/2$ quantum Hall state\cite{MR}. Another way to realize an extended majorana operator is to tune a 1D
chain of localized majorana particles to a critical point between two topologically distinct gapped phases. \cite{Kitaev-1}  

\begin{figure}[b!]
\label{fig:g1-g2}
\subfigure[]{
\includegraphics[width=.5\textwidth]{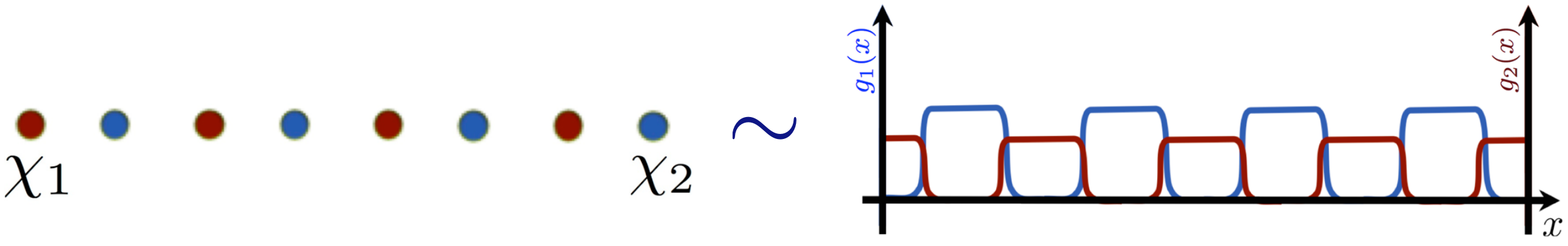}
}
\subfigure[]{
\includegraphics[width=.5\textwidth]{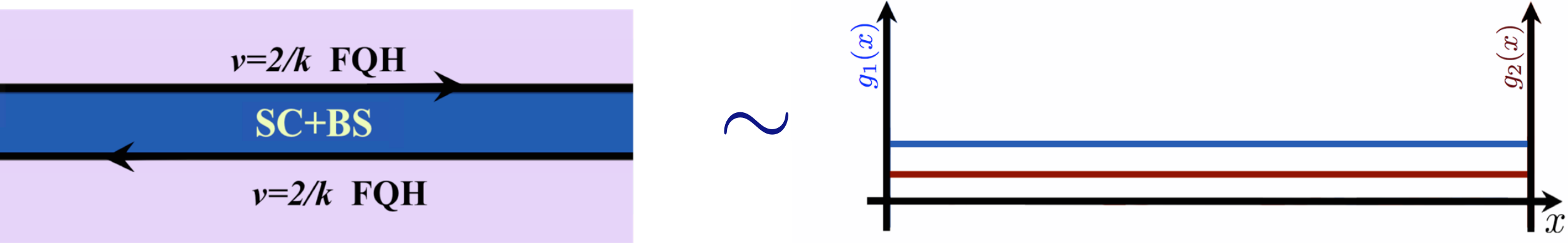}
}
\caption{Two distinct realizations of 1D parafermion systems. (a) A $\mathbb{Z}_k$ parafermion chain whose microscopic degrees of freedom are described by lattice parafermion operators $\chi_1$ and $\chi_2$. The low energy description of this system is given by a $\beta^2=2\pi k$ self-dual sine-Gordon model whose coupling constants $g_1$, and $g_2$ have the shown spatial profile. (b) Two counter propagating edge states of adjacent fractional quantum Hall states at $\nu=2/k$ filling fraction that are uniformly gapped through electron back-scattering (BS) and electron pairing (SC) perturbations. This system is equivalent to a $\beta^2=2\pi k$ self-dual sine-Gordon model with spatially uniform coupling constants $g_1$ (represents electron pairing) and $g_2$  (represents electron backscattering).}
\end{figure}

There is now growing interest in various ways of using hybrid structures to realize\cite{Mong-1,Vaezi-2} $\mathbb{Z}_k$ 
parafermions\cite{Fradkin-1}
 which are generalization of majorana fermions that amount to $\mathbb{Z}_2$ parafermions.  However parafermions other than Majorana remained illusive except for their possible realization through edge states of $\nu=12/5$ quantum Hall states.\cite{RR} Recently several proposals for engineering $\mathbb{Z}_3$ parafermions appeared\cite{PF-1,PF-2,PF-3, Vaezi-1, Bombin,Genon-B1,Toric-1, Genon-B2, Genon-Teo} and of our particular interest is the proposals for engineering two-dimensional (2D) $\mathbb{Z}_3$ parafermions using hybrid structures to build one-dimensional (1D) parafermion systems and coupling them\cite{Mong-1, Vaezi-2}. One approach sketched in Fig.~\ref{fig:g1-g2}(a) for achieving this goal is to start with chains of {\it localized parafermions} using a pair of spin-degenerate abelian fractional quantum Hall edge states and alternating host of back scattering centers (represented by finite $g_1$ regions in Fig.~\ref{fig:g1-g2}(a)) and superconducting islands (represented by finite $g_2$ regions  in Figure~\ref{fig:g1-g2}(a)). An alternate approach sketched in Fig.~\ref{fig:g1-g2}(b) to achieving this goal is to introduce uniform electron backscattering and pairing.  Both approaches yield gapped phases at generic electron backscattering and pairing strengths. However, when the pairing strength and electron backscattering strengths either of alternating setup for parafermion chain or of uniform setup are tuned to be equal, the system becomes critical\cite{Mong-1, Vaezi-2}. While the low energy effective theory of the quantum critical point (phase) should be identical for both approaches, i.e. the same conformal field theory (CFT), the relation between primary fields of the CFT and the short distance/ high energy degrees of freedom can be different. Such differences would impact coupling between so engineered 1D systems when attempting to build a 2D superconducting analogue of $\nu=12/5$ through coupling the 1D systems tuned to be critical.  

In this paper we carefully examine both cases focusing on deriving the relation between short distance, ultraviolet (UV) degrees of freedom that would couple 1D systems and primary fields of relevant low energy infrared (IR) CFT describing the critical point (phase) of the 1D systems. For the parafermion chain (see Fig.~\ref{fig:g1-g2}(a)) we accomplish this by first using the known mapping between the $\mathbb{Z}_k$ parafermion chain and the $\mathbb{Z}_k$ quantum clock chain but then bosonizing the quantum clock operators explicitly. This is distinct from the approach of projection employed in Ref.~\cite{Mong-1} for deriving the parafermionic nature of the localized domain wall states in the setup proposed for realizing parafermion chain. For the $\mathbb{Z}_k$ parafermion chain the fundamental UV degrees of freedom are localized parafermions. For the uniform backscattering and pairing approach (see Fig.~\ref{fig:g1-g2}(b)), we establish the relation between the UV degrees of freedom and the IR degrees of freedom through UV-IR transmutation of the self-dual sine-Gordon (SDSG) model. In this setting the fundamental UV degrees of freedom are the chiral anyons $e^{i\phi_R}$, $e^{i\phi_L}$.

The rest of the paper is organized as follows. In Section~\ref{sec:chain} we derive the low energy effective field theory of $\mathbb{Z}_k$ parafermion chain near the critical point or phase between two topologically distinct\cite{Fendley-1} gapped phases and obtain the relation between the domain wall parafermions and the primary fields of the relevant CFT's.
In Section~\ref{sec:uniform} we present the UV-IR transmutation of the SDSG model relevant for the limit of uniform back scattering and pairing tuned to equal strength. We conclude in Section~\ref{sec:conclusion} with discussions on implication of our results for describing coupled 1D systems among other things.  

\section{Critical point or phase of parafermion chain}\label{sec:chain}
The  $\mathbb{Z}_k$  parafermion chain of interest is a generalization of the Majorana chain made of fermion operators which are their own Hermitian conjugate.\cite{Kitaev-1} As the Hilbert space for a single Majorana operator is not well-defined, two flavors of Majorana operators per lattice site were used to construct the model of Majorana chain in Ref. \onlinecite{Kitaev-1}. We study a generalization of the Majorana chain consisting of  
two flavors of  $\mathbb{Z}_k$ parafermion operators $\chi_{1,i}$ and $\chi_{2,i}$ at site $i$ defined by the following algebra\cite{Fradkin-1,Nussinov-1,Fendley-1}
\begin{eqnarray} 
&&\chi_{a,i}^{k}=1,\quad \chi_{a,i}^{\dag}=\chi_{a,i}^{k-1},\quad \chi_{1,i}\chi_{2,i}=\omega \chi_{2,i}\chi_{1,i},
\cr
&&\chi_{a,i}\chi_{b,j}=\omega \chi_{b,i}\chi_{a,j}, \quad\quad i<j,\quad \quad a,b=1,2.
\label{eq:para-def}
\end{eqnarray}
in which $\omega=\exp\para{2\pi i/k}$. As $\para{e^{i\pi/k} \chi_{1,i}^\dag \chi_{2,i}}^{k}=1$ leads to $k$ distinct eigenvalues for $\chi_{1,i}^\dag \chi_{2,i}$ operators the lattice parafermions defined in  Eq.~\eqref{eq:para-def} span a $k$-dimensional local Hilbert space. Now we consider an infinitely long chain of $\mathbb{Z}_k$ parafermions with the following Hamiltonian
\begin{eqnarray}\label{para-1}
&&H^{\rm PF}_{k}=-\sum_{\alpha=1}^{k-1}\sum_{i}\left[J_{\alpha}\para{\bar{\omega}\chi_{2,i}^\dag \chi_{1,i+1}}^{\alpha}+h_{\alpha}\para{ \bar{\omega}\chi_{2,i}^\dag \chi_{1,i}}^{\alpha}\right]~~~
\end{eqnarray}
in which $\bar{\omega}=\omega^{-1}=\exp\para{-2\pi i/k}$, $J_{\alpha}=J^*_{k-\alpha}$, and $h_{\alpha}=h^*_{k-\alpha}$ for $\alpha=1,\cdots, k-1$. In this paper, we assume $J_{1}=J_{k-1}^*$, and $h_{1}=h_{k-1}^*$ are much larger than other coupling constants when $k>4$.\cite{Comment-0} For $k=2$ the above Hamiltonian reduces to the Kitaev Hamiltonian of Majorana chain.\cite{Kitaev-1} This Hamiltonian was studied as a generalization of the Majorana chain \cite{Kitaev-1} focusing on the edge state\cite{Fendley-1} or the Fock representation of the parafermions\cite{Cobanera} and Mong et.al.\cite{Mong-1} considered a system of coupled parafermion chains. 
We focus on bosonizing the parafermion chain to obtain the low-energy effective field theory and access the CFT associated with critical point or critical phase of the parafermion chain.

As an intermediate step of bosonizing the parafermion chain, we first employ the well-known mapping between the parafermion chain and $1+1$ dimensional quantum clock model 
.\cite{Fradkin-1,Nussinov-1,Fendley-1} 
The quantum clock model is defined in terms of two local operators $\sigma_i$ and $\tau_i$  satisfying the following commutation relations 
\begin{eqnarray}\label{Com-1}
&&\sigma_{i}^{k}=\tau_{i}^{k}=1,\quad \sigma_{i}^\dag =\sigma_{i}^{k-1},\quad \tau_{i}^\dag=\tau_{i}^{k-1},\cr
&&\sigma_{i}\tau_{i}=\omega \tau_{i}\sigma_{i}, \quad \sigma_{i}\tau_{j}=\tau_{j}\sigma_{i} \quad i\neq j,
\end{eqnarray}
and acting on the local Hilbert space of $k$ distinct states $\ket{m}_i$ with $m=1,\cdots,k$ through $\sigma_i\ket{m}_i=\omega^m\ket{m}_i$ and $\tau_i\ket{m}_i=\ket{m+1}_i$. 
The parafermion chain can now be mapped into the quantum clock model through expressing the $\mathbb{Z}_k$ parafermion operators in terms of the operators of Eq.~\eqref{Com-1} as
\begin{eqnarray} \label{Chi-1}
&&\chi_{1,i}=\sigma_{i}\prod_{l<i}\tau_{l}, \quad \chi_{2,i}=\omega \tau_{i}\chi_{1,i}
\end{eqnarray}
which satisfy the algebra of Eq.~\eqref{eq:para-def}.
Now the parafermion chain Hamiltonian of Eq.~\eqref{para-1} becomes
\begin{eqnarray}\label{Clock-1}
&& H_{k}=-\sum_{\alpha=1}^{k-1}\sum_{i}\left[J_{\alpha} \para{\sigma_{i+1}^\dag \sigma_{i}}^{\alpha}+h_{\alpha}\tau_{i}^{\alpha}\right],
\end{eqnarray}
which is the Hamiltonian of $\mathbb{Z}_k$ clock model. 

The $\mathbb{Z}_k$ clock model Hamiltonian clearly shows that there are two gapped phases in the model: ferromagnetic and paramagnetic phases. A convenient way to access the quantum critical point is through use of duality. The duality transformation of the model Eq.~\eqref{Clock-1} can be obtained by considering the following operators:
\begin{eqnarray}\label{Com-2}
&& \mu_{i}=\prod_{l<i}\tau_{l}, \quad \nu_{i}=\sigma_{i}^\dag \sigma_{i+1}.
\end{eqnarray}
that satisfy the same commutation relations as the original clock operators in Eq. \eqref{Com-1}. In terms of the new clock operators, the Hamiltonian in Eq. (\ref{Clock-1}) reads
\begin{eqnarray}\label{Clock-2}
&& {H}^{\rm dual}_{k}=-\sum_{\alpha=1}^{k-1}\sum_{i}\left[J_{\alpha}\nu_{i}^{\alpha} +h_{\alpha}\para{\mu^\dag_{i+1}\mu_{i}}^{\alpha}\right],
\end{eqnarray}
which is related to the Hamiltonian in Eq. \eqref{Clock-1} through $\sigma_i \leftrightarrow  \mu_i$, $\tau_i\leftrightarrow \nu_i $, and $J_{\alpha} \leftrightarrow h_{\alpha}$ duality transformations. Thus, $J_{\alpha}=h_{\alpha}$ (for every $\alpha$) is the self-dual point and hence a critical point of the clock model, provided the phase transition is of the second order.

\begin{figure}[t]\label{fig:PD}
\includegraphics[width=.4 \textwidth]{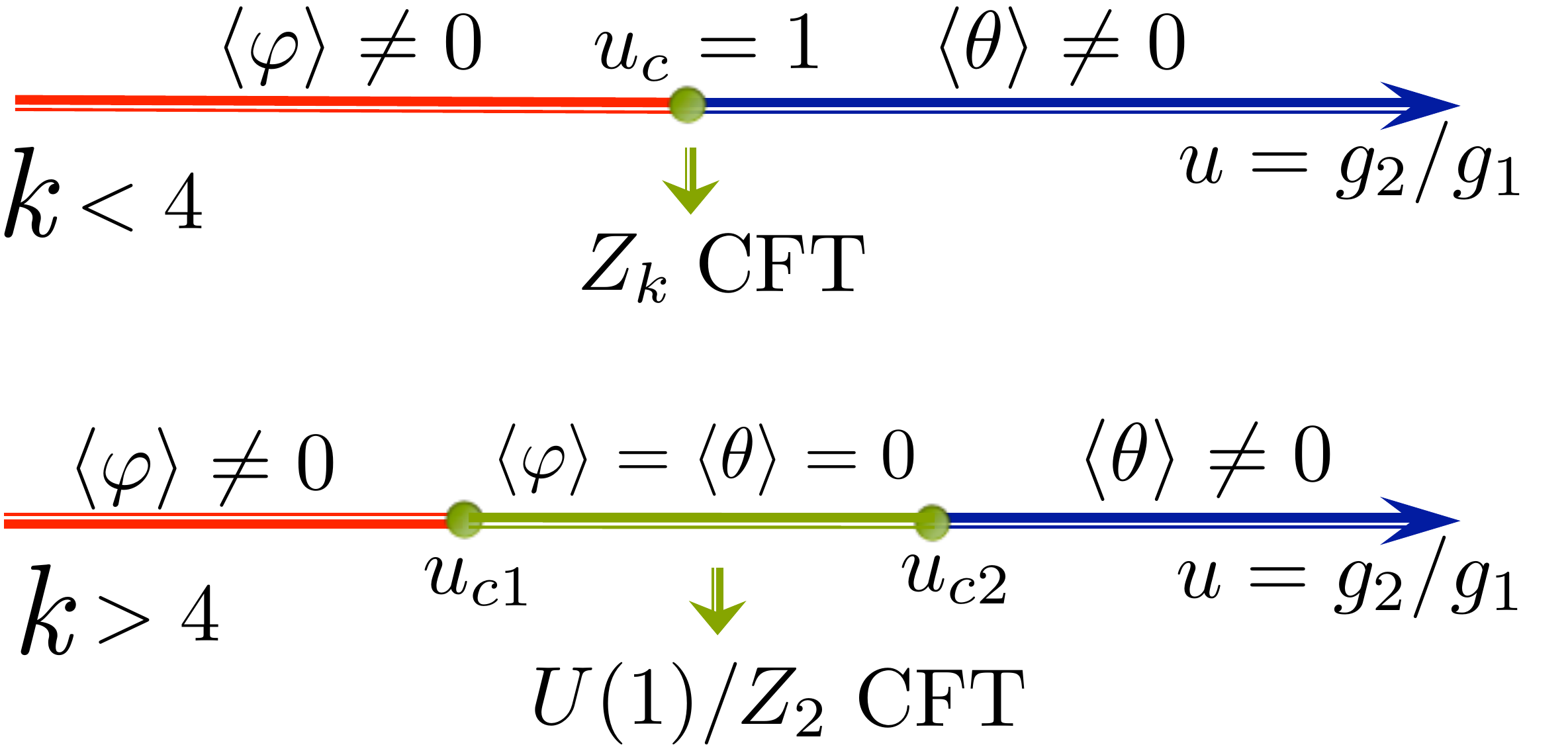}
\caption{The zero temperature phase diagram of  ``$\beta^2=2\pi k$ SDSG model'' as a function of a single tuning parameter $u\equiv g_2/g_1$. Two gapped phases are associated with condensation of $\varphi$ and $\theta$ respectively.} 
\end{figure}

For $k>4$, the phase diagram for the model Eq.~\eqref{Clock-1} has been known from classical statistical mechanics of two-dimensional clock model to be rather rich with 
 the critical phase for $k>4$ when $J_1\gg J_\alpha$, for $\alpha\neq1$\cite{CM-1,CM-2} (see Fig.~\ref{fig:PD}). Though it has not been explicitly stated in the literature, the observations of Refs.~\onlinecite{Gehlen, Suranyi} predating establishment of orbifold CFT\cite{Orbifold-1,Orbifold-2} strongly suggests that the CFT associated with the critical phase is $U(1)_k/\mathbb{Z}_2$. In this case the primary fields of the CFT are free bosons and the lattice parafermion operators are trivially related to the vertex operators of the bosons. The situation is quite different for $k<4$. Though it has been known 
 that the critical point is described by $\mathbb{Z}_k$ parafermion CFT\cite{ZF-1} for $k<4$, relating the lattice parafermion operators of a single parafermion chain to the primary fields of  $\mathbb{Z}_k$ parafermion CFT is less trivial. To date such relationship has only been conjectured\cite{Mong-1}.

Now we bosonize the clock model of Eq.~\eqref{Clock-1} to derive the explicit relationship between the lattice parafermion operators of a single parafermion chain and the primary fields of the $\mathbb{Z}_k$ parafermion CFT for $k<4$. For this we need to find all symmetries associated with the Hamiltonian and the associated charge of operators. The clock model in Eq. \eqref{Clock-1} enjoys the following global $\mathbb{Z}_k$ symmetry: $\sigma_{i}\to \omega \sigma_{i}$, and $\tau_{i}\to \tau_{i}$. Similarly, the dual Hamiltonian in Eq. \eqref{Clock-2} is symmetric under a different $\mathbb{Z}_k$ transformation: $\mu_{i}\to \omega \mu_{i}$, and $\nu_{i}\to \nu_{i}$. We call this latter symmetry $\mathbb{Z}^{\rm dual}_k$ and the corresponding charge of operators the {\em dual $\mathbb{Z}_k$ charge}. Hence, $\tau_i$, and $\nu_i$ are invariant under both symmetry transformation, while $\sigma_i$ and $\mu_i$ carry $(1,0)$ and $(0,1)$ charges under $\mathbb{Z}_k$ and $\mathbb{Z}_k^{\rm dual}$ symmetries, respectively.

Knowing the charge of clock operators, we are in a good position to to bosonize the clock model. To this end, we first define two conjugate bosonic fields $\varphi$ and $\theta$ with
\begin{eqnarray}\label{com-4}
&&\left[\varphi\para{x},\varphi\para{x'}\right]=\left[\theta\para{x},\theta\para{x'}\right]=0,\cr
&&\left[\varphi\para{x},\theta\para{x'}\right]=i\frac{\pi}{k}{\rm sgn}\para{x-x'},
\end{eqnarray}
commutation relations and
\begin{eqnarray}\label{trans-1}
&&\mathbb{Z}_{k}^{~~~~}: \quad \varphi \to \varphi +2\pi/k,\quad \theta\to \theta,\cr
&&\mathbb{Z}_{k}^{\rm dual}: \quad \varphi\to \varphi,\quad\quad \quad \quad \theta\to \theta+2\pi/k.
\end{eqnarray}
symmetry transformation properties. Since $\sigma_i$, $e^{i\varphi}$, and $e^{-i\para{k-1}\varphi}$ are all neutral under $\mathbb{Z}_{k}^{\rm dual}$ and carry a unit charge of $\mathbb{Z}_k$ symmetry, it is tempting to write
\begin{eqnarray}\label{Sigma-1}
&& \sigma_{i}\sim  \exp\para{i\varphi(i)}+a_{1} \exp\para{-i\para{k-1}\varphi(i)}+\cdots,
\end{eqnarray}
where $a_1$ is a non-universal constant. Similarly, $e^{i\theta}$ and $e^{-i\para{k-1}\theta}$ both carry the same charge as $\mu_i$, we conjecture the following identity
\begin{eqnarray}\label{Mu-1}
&& \mu_{i}\sim \exp\para{i\theta(i)}+b_{1}\exp\para{-i\para{k-1}\theta(i)}+\cdots~.
\end{eqnarray}
Again $b_{1}$ is a non-universal constant. 
Using the definition of $\tau_i$, and $\nu_i$ operators, at continuum limit we have:
\begin{eqnarray}\label{Tau-Nu}
&& \nu_{i}=\sigma_{i}^\dag \sigma_{i+1} \sim 1+ic_1\partial_x \varphi+c_2\para{i\partial_x \varphi}^2+c_3 \cos\para{k\varphi}+\cdots\cr
&& \tau_{i}=\mu_{i}^{\dag}\mu_{i+1}  \sim 1+id_1\partial_x \theta+d_2\para{i\partial_x \theta}^2+d_3 \cos\para{k\theta}+ \cdots,~~~
\end{eqnarray}
where $\cdots$ denotes the less relevant terms $c_l$'s and $d_l$'s for $l=1,2,3$ are again non-universal constants which depend on microscopic details. 
Since $e^{i\varphi\para{x'}}i\partial_x \theta\para{x}e^{-i\varphi\para{x'}}=i\partial_{x}\theta\para{x}-2\pi/k\delta\para{x-x'}$, it is straightforward to verify that Eqs. \eqref{Sigma-1}, \eqref{Mu-1}, and \eqref{Tau-Nu} give rise to the right commutation relations in Eq. \eqref{Com-1}. 

Having successfully bosonized $\sigma_{i}^\dag \sigma_{i+1}$, and $\tau_{i}$ we are prepared to obtain the low energy effective Hamiltonian of the clock model and of the parafermion chain. Inserting Eq.~\eqref{Tau-Nu} into  Eq.~\eqref{Clock-1} we arrive at
\begin{eqnarray}\label{SG-1}
\mathcal{H}_{SG,k}=&&\frac{k}{4\pi}\int dx ~v_{\rm F} \para{K\para{\partial_x \varphi}^2+K^{-1}\para{\partial_x \theta}^2}\cr
-&&\int dx~g_1\cos\para{k\varphi}+g_2\cos\para{k\theta}+\cdots ,
\end{eqnarray}
where we have absolved non-universal constants into the Fermi velocity $v_{\rm F}$, Luttinger parameters $K$, and renormalized coupling constants $g_1$ and $g_2$.  If we neglect the less relevant terms in Eq. \eqref{SG-1} denoted by $\cdots$, we obtain the so called ``$\beta^2=2\pi k$ SDSG model'' as the low energy description of the quantum clock model, and the parafermion chain. Note that it is crucial to include both terms with the same charges in Eqs. \eqref{Sigma-1} and \eqref{Mu-1} in order to arrive at the SDSG model with the $\mathbb{Z}_k$ symmetry required for the low-energy effective theory of the quantum clock model and the parafermion chain. This SDSG model is self-dual under a duality transformation: $\varphi \leftrightarrow \theta$, $g_1 \leftrightarrow g_2$, and $K\leftrightarrow K^{-1}$. Thus the self-dual point of the quantum clock model $\forall\alpha: J_\alpha=h_\alpha$ maps to $g_1=g_2$ and $K=1$. However, through this derivation we see that $g_1=g_2=0$ can happen only when all the microscopic parameters vanish which would imply $v_F=0$ in that case as well. Therefore, the Fermi velocity is not independent of $g_1$ and $g_2$ in this construction.

The phase diagram for the classical two dimensional $\mathbb{Z}_k$ clock model\cite{CM-2} implies the phase diagram sketched in Fig.~\ref{fig:PD} at zero temperature for the SDSG model in Eq. \eqref{SG-1} as a function of . For $k<4$, there is a single critical point described by $\mathbb{Z}_k$ parafermion CFT.\cite{Lech-1,int-1} For $k>4$ there exist a gapless region around the self-dual point described by $U(1)_k/\mathbb{Z}_2$ orbifold CFT\cite{Suranyi,Gehlen}. At $k=4$, there is a single critical point with $U(1)/\mathbb{Z}_2$ orbifold CFT description, but the boson compactification radius is non-universal\cite{Lech-1}. In the rest if this section we will focus on relating the lattice degrees of freedom to the primary fields of $\mathbb{Z}_k$ parafermion CFT describing the critical point for $k<4$.

For $k=2$, the SDSG model or the Majorana chain can be solved exactly\cite{MF-1,Lech-1}. For generic values of $g_1$, and $g_2$, the model can be mapped to two massive Majorana fermions with $g_1-g_2$, and $g_1+g_2$ masses. At the $g_1=g_2$ self-dual point, one Majorana modes becomes massless. Another approach is to start with two coupled Ising chains (Ashkin-Teller model) and then bosonizing them\cite{MF-2}. 
These two approaches suggest that the lattice Majorana operators $\chi_{1}$, and $\chi_{2}$ transmute to $\psi_{R}+\psi_{L}$, and $\psi_{R}-\psi_{L}$ primary fields of the Ising CFT, respectively.

Relating ``lattice'' operators in the $\mathbb{Z}_3$ quantum clock model and parafermion chain to the primary fields of $\mathbb{Z}_3$ parafermion CFT is less trivial as no exact solution is available. The IR fixed point  associated with finite $g_1=g_2$ at the self-dual point is described by a $\mathbb{Z}_3$ parafermion CFT with $c=4/5$ central charge. The holomorphic (chiral) sector of this CFT contains six primary fields: identity $\mathbb{I}_{R}$, spin field $\sigma_{R}$ and its conjugate $\sigma_{R}^\dag$ with scaling dimensions $1/15$, parafermion field $\psi_{R}$ and its conjugate $\psi_{R}^{\dag}$ with scaling dimensions $2/3$, and the thermal operator, $\epsilon_{R}$, (also know as energy operator) with scaling dimension 2/5. The anti-holomorphic sector is identical. Note the subscripts $R$ or $L$ represent the operators being functions only of $z\equiv t+ix$ and hence holomorphic (chiral) or functions only of $\bar{z}\equiv t-ix$ and hence anti-holomorphic(anti-chiral) respectively. In order to relate the ``lattice'' parafermions to primary fields of $\mathbb{Z}_3$ CFT, we use the fusion rules of the $\mathbb{Z}_3$ CFT and the charges of the fields under symmetry operations. 

In addition to the $\mathbb{Z}_3\times \mathbb{Z}_3^{\rm dual}$ symmetry of the clock model which already played an important role in the bosonization of parafermion chain, the 
charge of the operators under the symmetry operations on the chiral sectors guide our 
derivation of the relation between the ``lattice'' parafermions and the primary fields of $\mathbb{Z}_3$ CFT.
For this purpose we rewrite the SDSG model in terms of chiral, $\phi_{R}$, and anti-chiral, $\phi_{L}$, fields defined by $\phi_{R/L}\equiv\varphi \pm \theta $. Note that for $k=3$ the SDSG model is invariant under 
\begin{equation}
\mathbb{Z}_{3}^{R/L}:~\phi_{R/L}\to \phi_{R/L}+4\pi/3.
\label{eq:Z3RL}
\end{equation}
Now we can relate the charge of an operator under $\mathbb{Z}_3 \times \mathbb{Z}_3^{\rm dual}$ symmetry $(q,q^*)$ to its charge under $\mathbb{Z}^{R}_3 \times \mathbb{Z}_3^{L}$ as 
\begin{equation}
(q^R,q^L)=(q+q^*,q-q^*). 
\label{eq:qq*}
\end{equation}
The above equation implies that for chiral operators (with $q^L=0$), we have $q=q^*$, and for anti-chiral operators $q=-q^*$. The fusion rules of the $\mathbb{Z}_3$ parafermion CFT~\cite{ZF-1,RMP-1} suggest that $\sigma_{R}$, $\psi_{R}$, and $\epsilon_{R}$ chiral primaries carry $q^R=\{1, 2,3\}$ charges under $\mathbb{Z}_{3}^{R}$ respectively. Anti-chiral primary fields also carry the same charges but under $\mathbb{Z}_{3}^{L}$. Using Eq. \eqref{eq:qq*}, we can also find $(q,q^*)$ for the primary fields (see Table~\ref{Tab1}).

Now we use the charge of the operators under all the discrete symmetries we discussed so far combined with the fusion rules of the $\mathbb{Z}_3$ CFT to express the operators defined on the chain to primary fields of $\mathbb{Z}_3$ CFT. 
The lattice parafermion operators $\chi_{1,i}$, and $\chi_{2,i}$ without center of mass momentum and assigned sense of chirality 
cannot be simply related to the parafermion primary fields $\psi_{R/L}$ of $\mathbb{Z}_3$ CFT which are chiral.
$\chi_{1,i}$, and $\chi_{2,i}$  cannot be related 
to $\psi_R\psi_L$ either, since this composite object carries different charges
under symmetry operations. 
To relate the lattice parafermion operators $\chi_1$, and $\chi_2$ 
 to the primary fields of the $\mathbb{Z}_3$ parafermion CFT, we first note their relation to quantum clock model operators:  $\chi_{2,i} = \sigma_i\mu_i$ and $\chi_{1,i}=\bar{\omega}\tau^{\dag}_i\chi_{1,i}$ (see Eq.~\eqref{Chi-1}).

Based on the fact that the $\sigma_i$ operator is a non-chiral operator with zero conformal spin, and it carries $(1,0)=(1/2,1/2)+(1/2,-1/2)$ charge under $\mathbb{Z}_3\times \mathbb{Z}_3^{\rm dual}$,  
we can relate it to a product of $\sigma_{R}$ and $\sigma_L$ primary fields each carrying $(1/2,1/2)$, and $(1/2,-1/2)$ charges respectively. 
Thus we assume $\sigma_i \rightarrow \sigma_{R}\para{z}\sigma_{L}\para{\bar{z}}$ in the continuum limit, where $\rightarrow$ indicates relating UV limit degrees of freedom to the left of the symbol to the IR limit degrees of freedom to the right of the symbol. Similarly, $\mu_i \rightarrow \sigma_{R}\para{z}\sigma_{L}^\dag\para{\bar{z}}$. 
This implies 
\begin{equation}
\chi_{2,i} \rightarrow \sigma_{R}\para{z}\times \sigma_{R}\para{z} \sigma_{L}\para{\bar{z}} \times \sigma_{L}^\dag\para{\bar{z}}.
\label{eq:chi-sigma}
\end{equation}
However, from the fusion rules of the $\mathbb{Z}_3$ parafermion CFT \cite{ZF-1,RMP-1}
$\sigma \times \sigma \sim \sigma^\dag + \psi^\dag$, and $\sigma \times \sigma^\dag \sim 1+\epsilon$. 
Hence we can rewrite Eq.~\eqref{eq:chi-sigma} as 
$\chi_{2,i}  \rightarrow l_1 \psi_R^\dag + l_2\sigma_R^\dag \epsilon_L + l_3\sigma_{R}^\dag+l_4\psi_{R}^\dag \epsilon_{L}$, where $l_a$ are non universal constants in general. However, it is simple to see that $l_3=l_4=0$ is required for the charges of the two sides to match
(noting the fact that $\chi_{2,i}$ carries charge (1,1), and $\epsilon_{L}$ carries charge (3/2,-3/2) under $\mathbb{Z}_3\times \mathbb{Z}_3^{\rm dual}$ operations). 
Going through the similar procedure for $\chi_{1,i}$ 
we arrive at the following transmutation for the lattice parafermion operators.

\begin{eqnarray} \label{Chi-3}
&&\chi_{2,i}\rightarrow l_{1}\psi_{R}^{\dag}+l_{2}\sigma_{R}^{\dag}\epsilon_{L}+\cdots,\cr
&&\chi_{1,i} \rightarrow l'_{1}\psi_{R}^{\dag}+l'_{2}\sigma_{R}^{\dag}\epsilon_{L}+\cdots.
\end{eqnarray}
The non-universal constants $l_{i}$, and $l'_{i}$ are such that the $\chi_{1}$, and $\chi_2$ parafermions become (linearly) independent operators. 
We could also define two other flavors of parafermions using $\sigma_{i}$, and $\mu_{i}$ through $\gamma_{2,i}=\sigma_{i}\mu_{i}^\dag$, $\gamma_{1,i}=\gamma_{2,i}\tau_{i}$. These new parafermions transmutate as
\begin{eqnarray} \label{Gamma-1}
&&\gamma_{2,i} \rightarrow l_{1}\psi_{L}^{\dag}+l_{2}\sigma_{L}^{\dag}\epsilon_{R}+\cdots~\cr
&&\gamma_{1,i} \rightarrow l'_{1}\psi_{L}^{\dag}+l'_{2}\sigma_{L}^{\dag}\epsilon_{R}+\cdots.
\end{eqnarray}
The above relations are consistent with the conjectures made in Ref. \onlinecite{Mong-1}. The transmutation of UV degrees of freedom of parafermion chain is summarized in the Table~\ref{Tab1}.

\begin{table}[htbp]
  \centering
  \begin{tabular}{@{} ccc @{}}
    \toprule
     UV limit & IR limit & $ \mathbb{Z}_3 \times \mathbb{Z}_3^{\rm dual}$ charges\\ 
   \hline 
    $\sigma_{i}$ & $\sigma_{R}\sigma_{L} $&(1,0)\\ 
    $\mu_{i}$ & $\sigma_{R}\sigma_{L}^\dag $&(0,1)\\ 
    $\chi_{1,i}$ & $l'_{1}\psi_{R}^{\dag}+l'_{2}\sigma_{R}^{\dag}\epsilon_{L}$& (1,1) \\ 
    $\chi_{2,i}$ & $l'_{1}\psi_{R}^{\dag}+l_{2}\sigma_{R}^{\dag}\epsilon_{L}$&(1,1)\\ 
$e^{i\phi_{R}/2}+h_{1}e^{-i\phi_{R}}+\cdots$& $\sigma_{R} $&(1/2,1/2)\\ 
    $e^{-i\phi_{R}}+p_{1} e^{2i\phi_{R}}+\cdots$& $\psi_{R} $&(-1,-1)\\ 
    $\sin\para{3\phi_{R}/2}+\cdots$ & $\epsilon_{R} $&(3/2,3/2)\\ 
    \hline 
  \end{tabular}
  \caption{A summary of the UV-IR transmutation. For a parafermion chain (see Fig.~\ref{fig:g1-g2}(a)), well defined operators in the UV limit are lattice parafermion operators or clock operators which  transmute to nontrivial combination of the primary fields of  $\mathbb{Z}_3$ parafermion CFT. On the other hand for the uniformly gapped quantum Hall edge state pairs(see Fig.~\ref{fig:g1-g2}(b)), the primary fields of the UV limit given by free boson CFT are chiral and anti-chiral vertex operators. Their transmutation is given in the last three rows.}
  \label{Tab1}
\end{table}

\section{Uniform backscattering and pairing case}\label{sec:uniform}
The proposal for uniformly gapping fractional quantum Hall states at filling factor $\nu=2/3$ through both electron back-scattering and pairing also leads to the SDSG model of Eq.~\ref{SG-1} as low energy effective theory like the case of the parafermion chain, but with an important difference:  the velocity $v_F$ and the values of $g_1$ and $g_2$ are not tied to same microscopic parameters. The speed is that of unperturbed $\nu=2/3$ edge states and $g_1$ and $g_2$ represent strengths of uniform backscattering and pairing, when the SDSG model of Eq.~\ref{SG-1} is used to describe the uniform backscattering and pairing case depicted in Fig.~\ref{fig:g1-g2}(b). By contrast as a low energy effective theory for the parafermion chain depicted in Fig.~\ref{fig:g1-g2}(a),relation between the effective parameters and the microscopic parameters of $J_\alpha$ and $h_\alpha$ implies that $v_F=0$ when $g_1=g_2=0$.

When $g_1$ and $g_2$ can both be set to be zero as in the case of uniform backscattering and pairing case, 
the first part of the SDSG Hamiltonian describes a free boson theory and the second part contains two different types of mass terms that do not commute with each other. Since the conformal (scaling) dimension of $g_1$ term is $k/2K$ and that of $g_2$ is $kK/2$ at the free boson CFT, both are relevant perturbations for $k<4$ at the fixed point $g_1=g_2=0$ (which leads to $K=1$) and $g_1$ and $g_2$ perturbations grow under the RG flow. Hence the free boson CFT associated with $g_1=g_2=0$ limit is the UV description.  
The critical point associated with $g_1=g_2\neq0$ is described by $\mathbb{Z}_k$ parafermion CFT with central charge $c=2(k-1)/(k+2)$  as opposed to the free boson CFT with central charge $c=1$.  
For $k=3$ the primary fields of the UV fixed theory are vertex operators i.e. $e^{i\para{m\phi_R+n\phi_L}/2}$ with $m, n\in \mathbb{Z}_{6}$, and current operators $\partial \phi_R$, and $\partial \phi_{L}$.

Relating the operators that are physical in the $UV$ limit to primary fields of the $IR$ fixed point theory is to identify the UV-IR transmutation. This problem for the SDSG model has been 
investigated by Delfino~\cite{Delfino} and Lecheminant et al. \cite{Lech-1} before, but their findings are not sufficient for our purposes. Specifically, the transmutation of the lattice parafermions we presented in section II was not discussed. Moreover, the transmutation of chiral fields (e.g., $\sigma_{R}$, or $\psi_{R}$) from the high energy bosonic fields were not determined. Here we revisit this problem for $k=3$ and utilize the charge of operators under the discrete symmetries of the model and the criticality of the self-dual point as our guiding principles.

Now we find UV-IR transmutation of primary fields of $\mathbb{Z}_3$ parafermion CFT. 
The spin order $\sigma$, and spin disorder, $\mu$, fields can be written as the product of its holomorphic and anti-holomorphic parts as $\sigma=\sigma_{R}\para{z}\sigma_{L}\para{\bar{z}}$, and $\mu=\mu_{R}\para{z}\mu_{L}\para{\bar{z}}$. Matching the charge of the both sides under $\mathbb{Z}^{R/L}_3$ 
\begin{eqnarray}\label{Order-1}
&& e^{i\phi_{R}/2}+h_{1}e^{-i\phi_{R}}+\cdots \rightarrow \sigma_{R}=\mu_{R},\cr
&& e^{i\phi_{L}/2}+h_{2}e^{-i\phi_{L}}+\cdots \rightarrow \sigma_{L}=\mu_{L}^\dag,~
\end{eqnarray}
where $h_{i}$ are non-universal constants and the arrows indicate transmutation between UV limit vertex operators to the left of the arrow and IR limit $\mathbb{Z}_3$ parafermion CFT primary fields to the right of the arrow.
Note that $\sigma_{R}$, and $\sigma_{L}$ are nonlocal operators, sine they carry fractional conformal spin that amounts to the branch cut in their correlation functions (with themselves). On the other hand, $\sigma$ and  $\mu$ operator carry zero conformal spin and are a local operators with respect to themselves, but are mutually non-local. However, the above transmutation of $\sigma_R/L$ combined with the fusion rule for two $\sigma$ operators
in the $\mathbb{Z}_3
$ parafermion CFT~\cite{ZF-1,RMP-1}:  $\sigma \times \sigma \sim \sigma^{\dag} +\psi_{R}^{\dag}\psi_{L}^{\dag}$ leads us to
\begin{eqnarray}\label{Psi-1}
\exp\para{-i\phi_{R/L}}+p_{1} \exp\para{2i\phi_{R/L}}+\cdots\rightarrow \psi_{R/L}.~
\end{eqnarray}
The UV-IR transmutation we obtained in Eq.~\eqref{Psi-1} is consistent with the information that $\psi_{R}$ ($ \psi_{L}$) carries $q=+1$ charge under $\mathbb{Z}_k$ and $q^{*}=1$ (-1) under $\mathbb{Z}_k^{\rm dual}$ symmetries.~\cite{ZF-1} 

To complete the UV-IR transmutation study, the last remaining primary field of  
the $\mathbb{Z}_3$ CFT is the thermal (energy) operator whose conformal dimension is $\para{h_{\epsilon},\bar{h}_{\epsilon}}=\para{2/5,2/5}$. It is known in the context of CFT that perturbing the critical point with thermal operator opens a mass gap in the spectrum of the $\mathbb{Z}_3$ parafermion CFT\cite{CM-2} . Now let us rewrite 
 the $k=3$ SDSG model Hamiltonian of Eq. \eqref{SG-1} into $\mathcal{H}_{Z_3}$ describing the SDSG model at its self dual point which is equivalent to $\mathbb{Z}_3$ parafermion CFT and deviation away from the self-dual point as~\cite{Comment-1}
\begin{equation}\label{eq:mass-perturb}
 \mathcal{H}_{SG,3}=
\mathcal{H}_{Z_3} + \int dx~\frac{g_1-g_2}{2}\para{\cos\para{3\varphi}-\cos\para{3\theta}}.
\end{equation}
Comparing Eq.~\eqref{eq:mass-perturb} with the fact that thermal operator opens a mass gap in the spectrum of the $\mathbb{Z}_3$ parafermion CFT\cite{CM-2} we obtain the UV-IR transmutation of the thermal operator as 
\begin{eqnarray}\label{Thermal-1}
 \frac{\cos\para{3\varphi}-\cos\para{3\theta}}{2}+\cdots \rightarrow \epsilon\para{z,\bar{z}}.
\end{eqnarray}
Note that the above prescription is consistent with two known facts about the thermal operator: firstly it is neutral under both $\mathbb{Z}_k$ and $\mathbb{Z}_k^{\rm dual}$ symmetries, secondly it is odd under duality transformation $\varphi \leftrightarrow \theta$. Finally we can easily read the holomorphic and anti-holomorphic parts of the thermal operator off the Eq. \eqref{Thermal-1} to find
\begin{equation}
\sin\para{3\phi_{R}/2}+\cdots \rightarrow \epsilon_{R}\para{z}\,~ \sin\para{3\phi_{L}/2}+\cdots\rightarrow \epsilon_{L}\para{\bar{z}}
\end{equation}

Table~\ref{Tab1} summarizes the relation between UV limit operators native to two distinct cases studied in this section and section~\ref{sec:chain}.  The distinction between the two cases is the difference in the physical degrees of freedom in the UV limit that can for instance couple between multiple 1D systems. While both the parafermion chain supported at the domain walls where $g_1$ and $g_2$ perturbation alternates in Fig.~\ref{fig:g1-g2}(a) and the uniform backscattering and pairing case in Fig.~\ref{fig:g1-g2}(b) flow to $\mathbb{Z}_3$ parafermion CFT at the critical point accessed by tuning either the width of the regions and maximum values of $g_1$ and $g_2$ to be the same or tuning $g_1=g_2$, the differences in the UV limit description requires different UV-IR transmutations. In particular, it is important to note that the lattice parafermion operators $\chi_{1,i}$ and $\chi_{2,i}$ are distinct from primary fields $\psi_R/L$ of the $\mathbb{Z}_3$ parafermion CFT. While a linear combination of vertex operators of free chiral boson representing the fractional quantum Hall edge quasi-particles transmute to $\psi_R/L$, the table ~\ref{Tab1} indicates that bosonic representation of lattice parafermion operators are more complex:
\begin{eqnarray} \label{Chi-2}
&&\chi_{2,i}=e^{i\phi_{R}}+f_{1} e^{-i2\phi_{R}}\cr
&& ~\sd\sd+e^{-i\phi_{R}/2}\left[f_{2}\sin\para{\frac{3\phi_{L}}{2}}+\cdots\right],\cr
&&\chi_{1,i}=\bar{\omega} \tau_{i}^\dag\chi_{2,i}=\bar{\omega}\para{1-id_{1}\partial_{x}\theta+...}\chi_{2,i},
\end{eqnarray}
where  $f_{i}$ are non-universal constants. The above result can also be directly achieved through combining Eqs. \eqref{Chi-1}, \eqref{Sigma-1}, \eqref{Mu-1}, and \eqref{Tau-Nu}.

\section{Conclusion}\label{sec:conclusion}
To summarize we examined two recently proposed ways of accessing $\mathbb{Z}_k$ parafermion CFT using Abelian quantum Hall edge states with the aid of engineered backscattering and proximity induced pairing sketched in Fig.~\ref{fig:g1-g2}, focusing on explicit derivation of the low energy effective theory describing critical point or phase.  Firstly we considered the $\mathbb{Z}_k$ parafermion chain which is a generalization of majorana chain, used its equivalence with $\mathbb{Z}_k$ clock model and derived $\beta^2=2\pi k$ SDSG model as low energy effective theory through bosonization of order and disorder operators of clock model. We pointed out that for $k>4$ the low energy effective theory for the critical phase separating two gapped phases (paramagnetic and ferromagnetic phases in the language of quantum clock model) is given by $U(1)_k/\mathbb{Z}_2$ orbifold CFT with central charge $c=1$. It has been known that the critical point  of the SDSG model for $k<4$ is described by $\mathbb{Z}_k$ parafermion CFT. However, we give an explicit derivation of a non-trivial relation between lattice parafermion degrees of freedom defined in the UV limit and the primary fields of $\mathbb{Z}_k$ parafermion CFT. Moreover we noted that the UV limit of the critical parafermion chain cannot be related to the UV limit of the SDSG model where the latter is a free boson theory. The reason being the relation between the parameters of the SDSG model and ``microscopic'' parameters of the parafermion chain dictates that the free boson part as well as the mass part of the SDSG Hamiltonian to vanish when tuned to $g_1=g_2=0$. 

Secondly we examined the SDSG model with independently controlled free boson part and the mass part of the Hamiltonian as a low energy effective Hamiltonian describing a pair of counter propagating $\nu=2/3$ fractional quantum Hall edge states that are uniformly gapped by backscattering term and the pairing term. This is a case where the free boson theory which is the UV limit of the SDSG model is physically accessible as a limit where both backscattering and pairing perturbations are turned off. Here we focused on the UV-IR transmutation between the primary fields of the UV limit CFT (free boson CFT) and the IR limit CFT($\mathbb{Z}_3$ parafermion CFT) at the critical point. Matching charge of operators under discrete symmetries of the SDSG model and using the known fusion rules of the two CFT's we derived UV-IR transmutation of all the primary fields of $\mathbb{Z}_3$ parafermion CFT when only the transmutation of the thermal operator had been previously discussed.\cite{Lech-1} 

Our results bare particularly important implications for effects of coupling in a 2D array of one-dimensional systems depicted in Fig.~\ref{fig:g1-g2}. When nearby parafermion chains $I$ and $J$ are coupled, a physical process would involve coupling between lattice parafermions for instance,  $\chi_{a,i}^{I}\gamma_{b,i'}^{J\dagger}$. The effect of such coupling in the IR limit would imply perturbations of form~\cite{Mong-1} $\lambda_1\psi^{I\dagger}_R\psi^J_L+\lambda_2 \sigma^{I\dagger}_{R}\sigma^{J}_{L}\epsilon^{I}_{L}\epsilon^{J}_{R}$ to the $\mathbb{Z}_3$ parafermion CFT describing each critical chains. However as the second term proportional to $\lambda_2$
involves all the gapless modes on $I$ and $J$ chains, such a perturbation will 
gap out all the parafermion modes unless $\lambda_1 \gg \lambda_2$.  On the other hand, when fractional quantum Hall edge states forming a uniformly gapped setting of Fig ~\ref{fig:g1-g2}(b) are coupled through fractional quantum Hall strip, a physical process would involve anyon backscattering written as $e^{im\para{\phi_{I,R}-\phi_{J,L}}}$, where $m=1,2,3$. The effect of such coupling in the IR limit would imply perturbations of form $\sigma^{I\dagger}_{R}\sigma^{J}_{L}$, $\psi^{I\dagger}_{R}\psi^J_{L}$, and $\epsilon^{I\dagger}_{R}\epsilon^{J}_{L}$ depending on the value of $m$. These perturbations gap out all the parafermion modes inside the system except the two outermost ones with opposite chiralities. Hence despite the two settings sharing the same IR limit CFT descriptions, a 2D array of uniformly gapped setting would require less fine tuning in reaching the Fibonacci phase~\cite{Mong-1, Vaezi-2} of the resulting 2D fractional topological superconductor formed of coupled 1D chains.\cite{Vaezi-1}

{\bf Acknowledgement}
We gratefully acknowledge useful discussions with E. Cobanera and G. Ortiz. This work was supported in part by NSF CAREER grant DMR-0955822 (A.V. and E.-A.K.) and in part by Bethe Postdoctoral Fellowship (A.V.).


\begin{thebibliography}{70}
\bibitem{Comment-0}
This assumption has been made to avoid the first order phase transition for for $k>4$. For example, when $J_1=J_{\alpha}$, and $h_1=h_{\alpha}$ for all values of $\alpha$, the model can be mapped to the (conventional) $k$-state Potts model that is known to have a first order phase transition for $k>4$.
\end{thebibliography}
\end{document}